\documentclass{elsart}

\usepackage[square]{natbib}
\RequirePackage{xspace}
\def\kaos{{\sc Kaos}$/\!${\small A1}\xspace}

\usepackage{graphicx}
\graphicspath{{figures/}}

\usepackage{amssymb}
\journal{Nucl. Instr. and Meth. in Phys. Res. A}

\begin{document}

\begin{frontmatter}

\title{Design criteria for multi-layered scintillating fibre arrays
    with inclined columns}

\author{P.~Achenbach\corauthref{cor}},
\ead{patrick@kph.uni-mainz.de}
\author{L.~Nungesser}, and
\author{J.~Pochodzalla}

\address{Institut f\"ur Kernphysik, Johannes Gutenberg-Universit\"at
    Mainz, Germany}

\corauth[cor]{Tel.: +49-6131-3925831; fax: +49-6131-3922964.}

\begin{abstract}
  Multi-layered scintillating fibre arrays read-out are commonly used
  as high resolution charged particle hodoscopes. Fibres of a
  ``column'' along the geometrical trajectory of incident particles
  are typically grouped to one pixel of a multi-channel read-out
  device. In some applications the incident particles will cross the
  detection plane with large angles w.r.t.\ the normal to the layers.
  Then, the packing of the fibres needs to be adapted to the incident
  particles and the columns need to be inclined. In this paper
  possible fibre array geometries are shown, relevant design criteria
  for detectors are discussed, and the effect of diverging particles
  incident on fibre arrays was studied using a Monte Carlo simulation.
\end{abstract}


\end{frontmatter}

\section{Introduction}
Detectors comprising scintillating fibres, packed together to form
arrays, and multi-channel photo-detectors have been used since the
1990s to track charged particles, see
e.g.~\cite{Baer1994,Bosi1996,Horikawa2004}. A fibre array is made from
layers of fibres, often with alternating layers off-set relative to
each other. The design of such detectors is governed by the
relatively small light yield of thin ($\oslash <$ 1\,mm) fibres.

Multi-channel photomultiplier tubes are especially suitable for fibre
read-out because of the good matching between the pixel size of the
photomultiplier and common fibre diameters, offering a significant
reduction in size and cabling with respect to conventional tubes.
Since the pioneering work of Kuroda~\cite{Kuroda1981} such tubes have
been developed in order to meet the demands on precise and reliable
tracking devices under
high-rate~\cite{FAROS1995,FAROS1996,Agoritsas1998}. In recent years
they have been continuously improved. The drawbacks of early devices
have been greatly reduced; modern multi-anode tubes exhibit little
cross-talk and a high gain uniformity between pixels. The Hamamatsu
Photonics R5900 series has been chosen by many experimental groups for
fibre read-out.

In modern experiments, fibre bundles involving a rather large number
of channels are easily read out via multi-channel photomultiplier
tubes, while the use of single-channel photomultiplier tubes is no
longer economical in terms of cost and space requirements.
Accordingly, multi-layered structures of packed scintillating fibres
coupled to multi-anode photomultiplier tubes became the preferred
choice for some fast detectors, the fibre trackers in the COMPASS
experiment~\cite{Horikawa1999} at CERN being one recent example.

Fibres of a ``column'' along the geometrical trajectory of incident
particles are typically grouped -- with one fibre from each layer ---
to one common pixel of the multi-channel read-out device.  If the
light yield per crossing particle is too small to be detected with the
required efficiency, layers need to be added to the array. In case
charged particles are crossing the fibre array at right angle to the
detector base the total thickness of the array can be increased
arbitrarily until a physical limit or restrictions in terms of small
angle scattering are reached. This condition applies to many
applications, as the angular acceptance for fibre hodoscopes is
generally small, typically up to $\theta=$ 3$^\circ$. However, in some
applications the incident particles will cross the detection plane
with large angles, $\theta$, w.r.t.\ the normal to the layers, and
would traverse several neighbouring columns.  The resultant large hit
multiplicities would compromise the tracking capabilities of such a
detector.  For the fibre hodoscopes of the DIRAC experiment the
detector response to particles with incident angles up to $\theta\sim$
45$^\circ$ was studied for background issues~\cite{Gorin2006}. The
read-out electronics of that detector lead to a strong suppression of
detection efficiency for larger angles.  For restoring the original
spatial resolution the packing of the fibres needs to be adapted to
the incident particles' direction and the columns need to be
inclined. Only the \kaos experiment is known in which the fibre column
angle, $\phi$, w.r.t.\ the normal to the layers is adapted to large
($\theta >$ 45$^\circ$) incident angles of the particles for
deliberately matching the
geometry~\cite{Achenbach-SNIC06,Achenbach-HYP06}. In the
spectrometer's electron arm hodoscope the average incident angle is
$\theta=$ 65$^\circ$. It has been shown that a high detection
efficiency and a good spatial resolution can be achieved using
detectors with inclined columns.

\section{Fibre array geometries with inclined columns}
\begin{figure}[htb]
  \centering
  \includegraphics[height=\textwidth,angle=90]{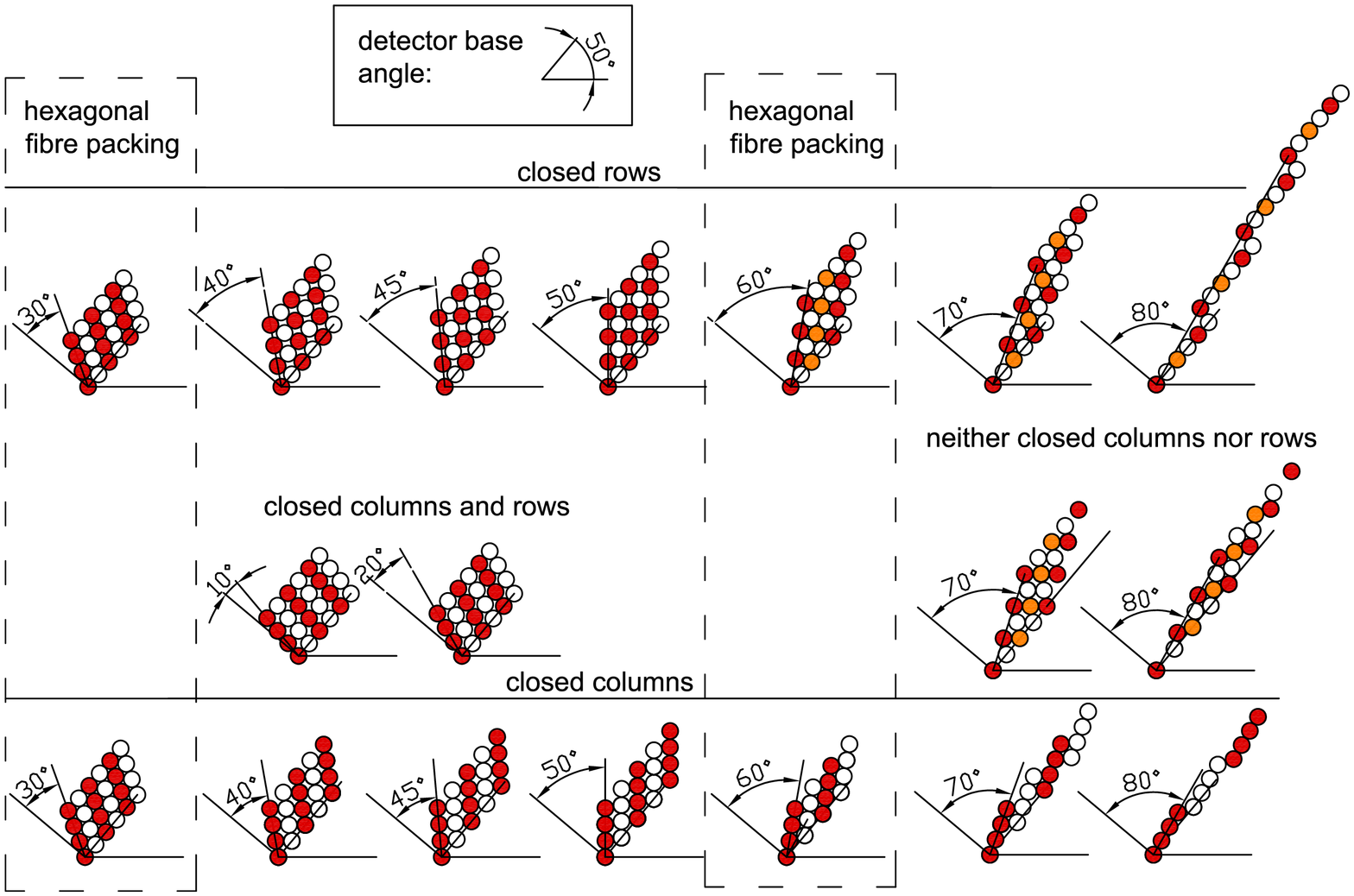}
  \caption{Different detector geometries with inclined columns (four
    fibres each) are shown. Column angles $\phi$ range from 10$^\circ$
    to 80$^\circ$ with a detector base angle of 50$^\circ$.
    Geometries with closed rows and closed columns are separately
    drawn. For column angles of 30$^\circ$ and 60$^\circ$ the fibre
    centres are forming a hexagonal lattice. For a column angle of
    45$^\circ$ the fibre centres are forming a square lattice.}
  \label{fig:geometries}
\end{figure}

Fibre array geometries with different column angles, $\phi$, ranging
from 10$^\circ$ to 80$^\circ$ and four fibres per column are shown
schematically in Fig.~\ref{fig:geometries}. In the scheme the base of
each detector is turned by an angle of 50$^\circ$ to the horizontal.
The most efficient way to pack fibres together to form inclined
columns is not obvious.  ``Rows'' of fibres are defined along the
layers in the direction of the detector base.  For column angles
between 30$^\circ$ and 60$^\circ$ the arrays can be formed in a way
that all fibres of a column touch each other (closed columns), but
leaving gaps to the corresponding fibres in neighbouring
columns. Alternatively, the arrays can be formed with corresponding
fibres of neighbouring columns touching (closed rows), leaving gaps
between fibres within a column.  Finally, for column angles $\phi$
below 30$^\circ$ both columns and rows can be closed.  For column
angles of 30$^\circ$ and 60$^\circ$ the fibre centres are forming a
hexagonal lattice. For a column angle of 45$^\circ$ the fibre centres
are forming a square lattice. The hexagonal packing, in which each
fibre is surrounded by 6 touching fibres, has the highest packing
density of $\pi/\sqrt{12}\simeq$ 0.90.  Square and hexagonal packing
geometry are used most often for fibre arrays.  Each geometry
corresponds to a different overlap fraction, a different column pitch,
a different detector width and length for a given number of layers and
read-out pixels, and a different average thickness.  The relevance of
these design criteria for choosing the detector geometry vary with the
specific application.  Primarily, the fibre array needs to provide a
high enough light yield per read-out pixel to discriminate the signals
against noise.  Light yield and detection efficiency depend on the
diameter and overlap fraction as well as on the number of layers.
Secondly, the spatial resolution depends on fibre diameter and column
pitch.  A detector geometry with closed columns as shown in the figure
will have a relatively large column pitch and a relatively small
spatial overlap. This leads to low spatial resolution and low
detection efficiency. Both can be avoided by placing the layers not
along the base of the detector, but with a smaller column pitch and
larger spatial overlap. That will create a shift between the first
fibres of each column and the base-line it was supposed to track. By
occasionally (every 3 or more columns) moving the column position one
full fibre diameter in the direction of the base-line the shift will
be corrected. These geometries have the advantage of being better
adapted to practical applications, but have less symmetry. For column
angles above 60$^\circ$ geometries with neither columns nor rows
closed can be created in a similar way.  In practice, only those fibre
arrays can be built which allow precise mounting and alignment during
all stages of the processing: gluing, bending, and installation.

\begin{figure}[htb]
  \centering
  \includegraphics[width=0.48\textwidth]{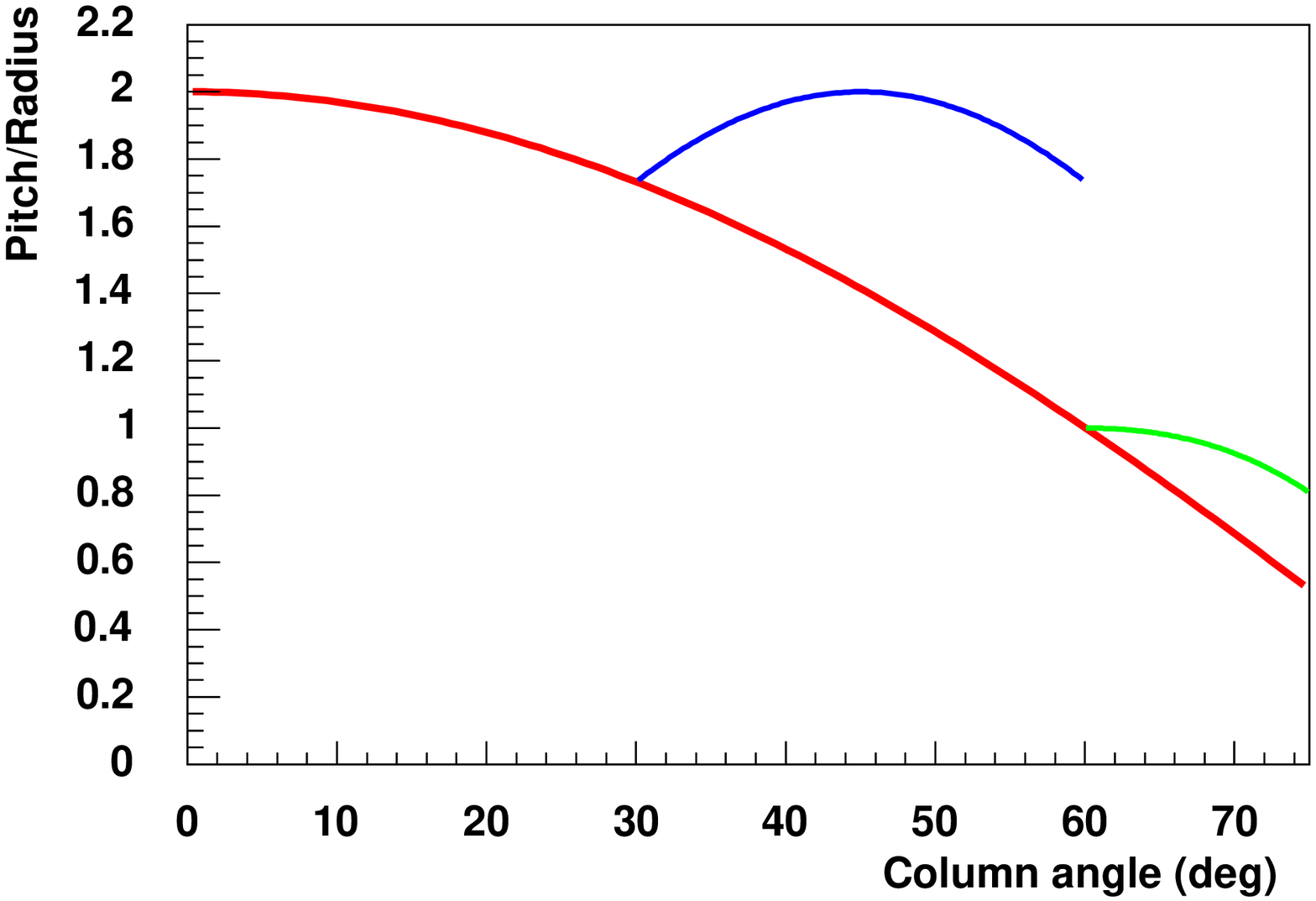}\hfill
  \includegraphics[width=0.48\textwidth]{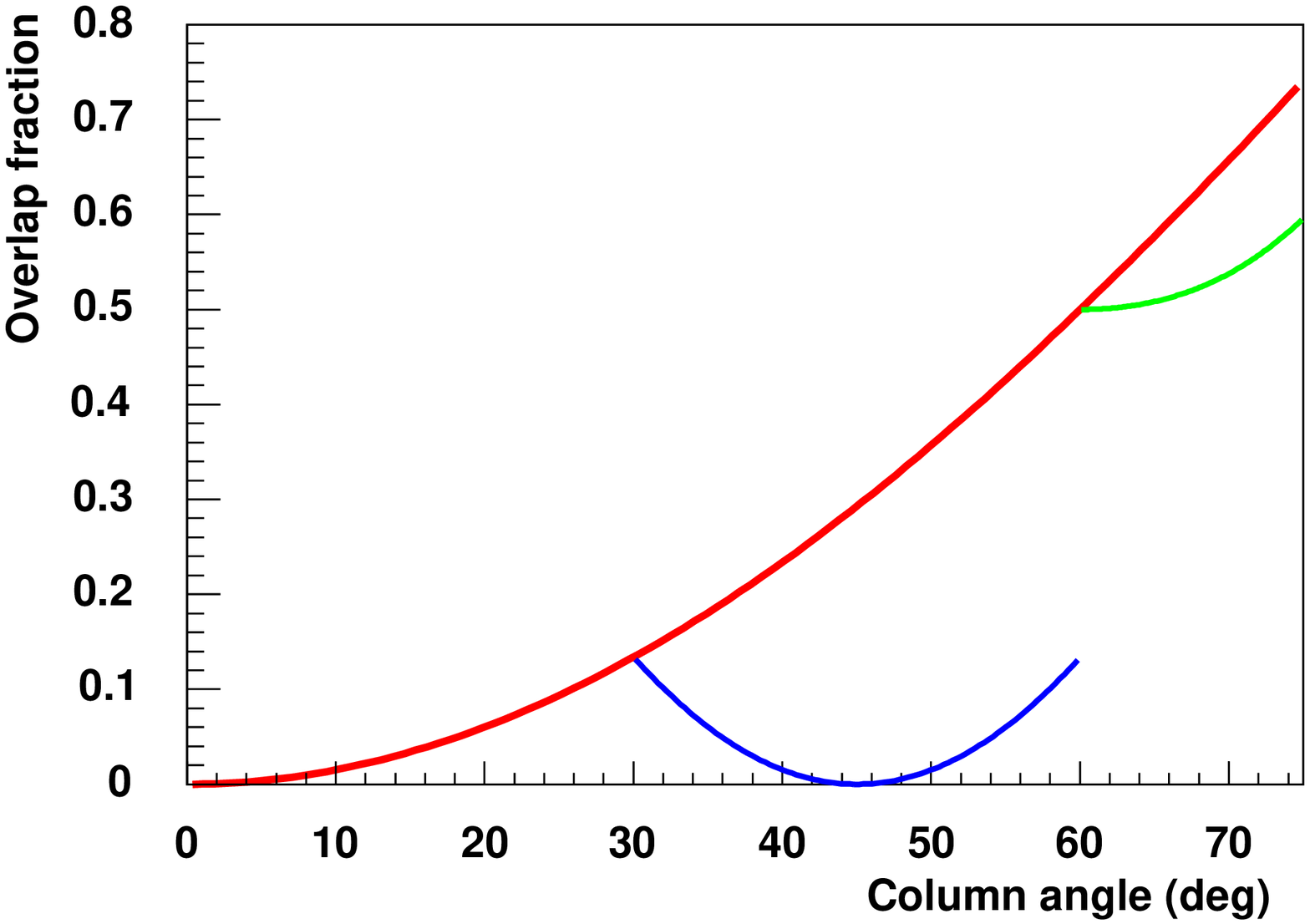}
  \caption{The left plot shows the ratio of the column pitch to the
    fibre radius as a function of the column angle $\phi$. The right
    plot shows the overlap of two neighbouring fibre columns in units
    of the fibre radius.  The continuous curve corresponds to a
    geometry with closed rows, the two branches that are splitting off
    to geometries with closed columns.}
  \label{fig:design}
\end{figure}

Fig.~\ref{fig:design}\,(left) shows the ratio of the column pitch to
the fibre radius, $p/\rho$, i.e.\ the distance between two fibre
centres along the base direction in units of fibre radius, as a
function of the column angle, $\phi$. The main branch starting at a
ratio $p/\rho=$ 2 corresponds to a geometry with both, closed rows and
closed columns, for which $p/\rho= 2\cos\phi$.  The branch splitting
off at $\phi= 30^\circ$ corresponds to a geometry with closed columns,
where $p/\rho= 4\sin\phi\cos\phi$. The continuous curve is calculated
for the geometry in which the rows are closed. The branch splitting
off at $\phi= 60^\circ$ follows a $p/\rho= 4\sin\phi / \sqrt{9 +
  \tan^2\phi}$ dependence and corresponds to closed columns.  Spatial
overlap is important to avoid relying on the detection of events with
only a grazing contact of the charged particle with the fibres. A
particle which crosses the gap between two fibres in one of the layers
needs to traverse a significant portion of the full fibre diameter in
the other layer. Fig.~\ref{fig:design}\,(right) shows the overlap
fraction, $o/\rho$, which is the overlap of two neighbouring fibre
columns in units of the fibre radius. It is directly related to the
pitch to radius ratio by $o/\rho= (2 - p/\rho)$.  Fibre diameter and
overlap fraction relate to the theoretical spatial resolution of fibre
arrays expected from the geometry. For events in which exclusively one
column was hit, the the spatial resolution is $\sigma= (\oslash -
2o)/\sqrt{12}$. For events in which the particle trajectory covered
purely the inter-column region, the spatial resolution is given by
$\sigma= o/\sqrt{12}$. Averaging both event types leads to a combined
resolution of $\sigma= \{(\oslash - 2o)^2 + o^2)/(\oslash -
o)\}/\sqrt{12}$. For fibre detectors in hexagonal packed arrays the
total overlap is $o= \oslash (1 - 1/\sqrt{2})$ with a pitch of $p=
\oslash/\sqrt{2}$. The theoretical spatial resolution of such an array
is $\sigma= \{9/\sqrt{2}-6\}\oslash/\sqrt{12}\approx 0.1\,\oslash$.

\begin{figure}[htb]
  \centering
  \includegraphics[width=0.48\textwidth]{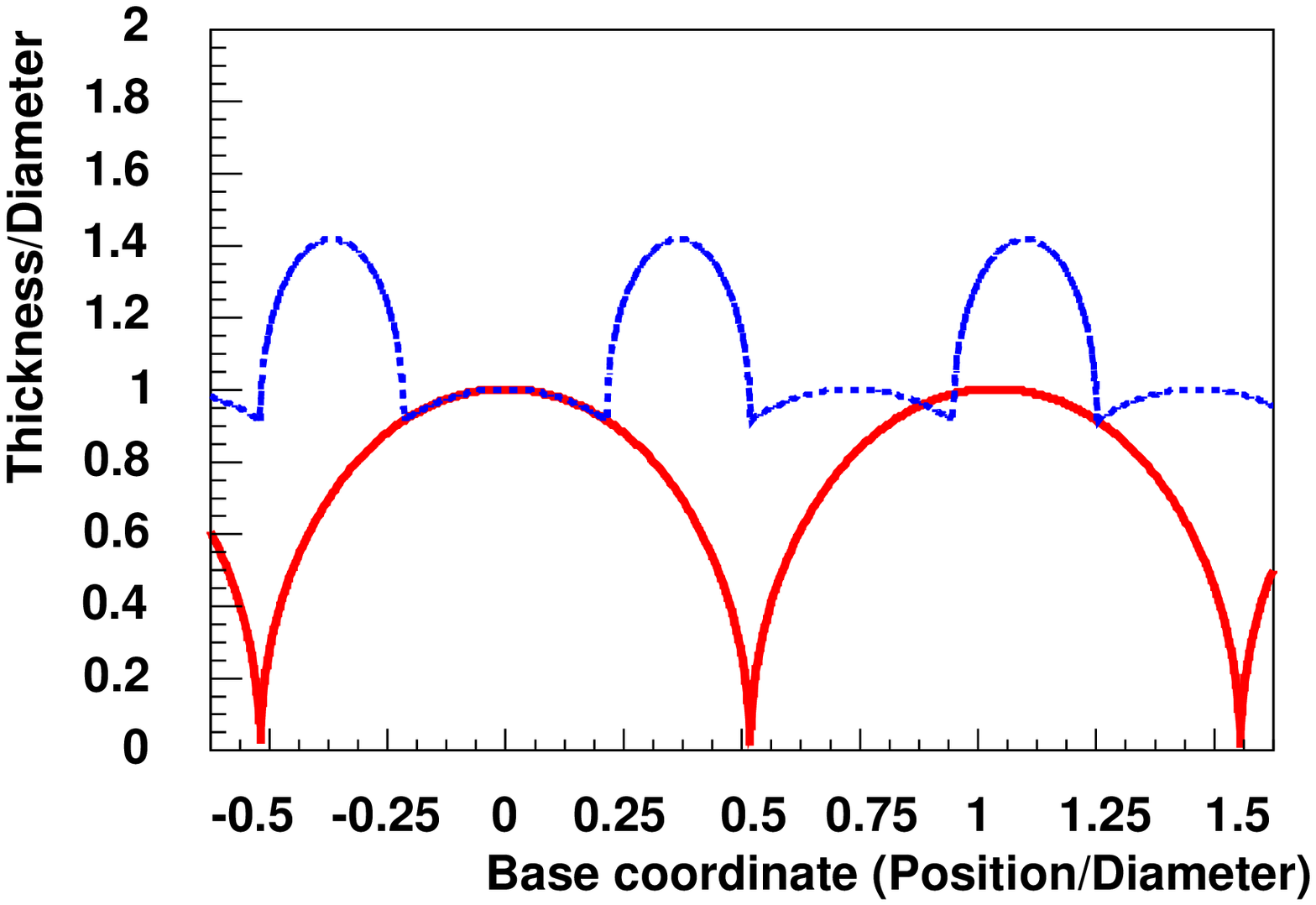}
  \includegraphics[width=0.48\textwidth]{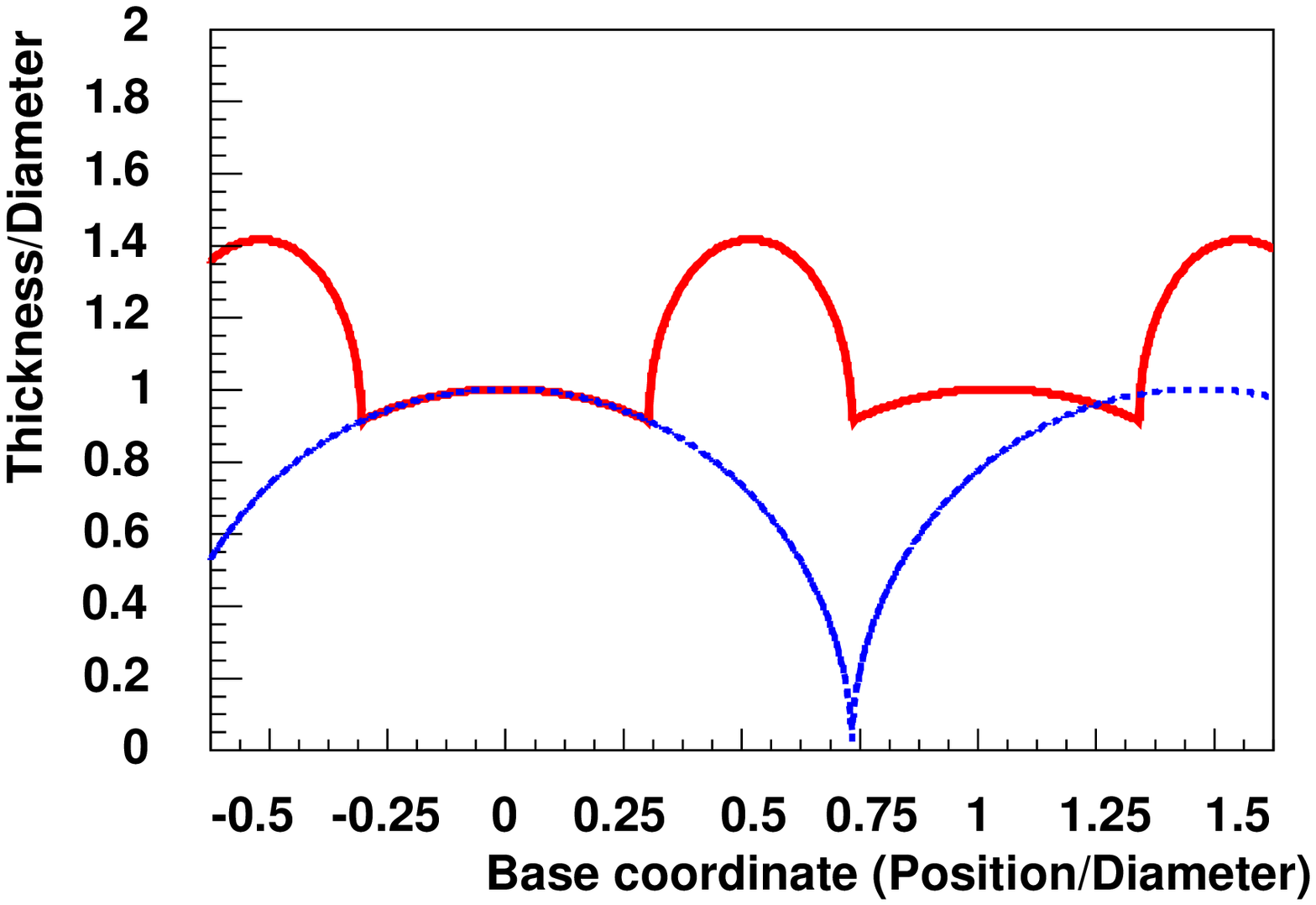}\\
  \includegraphics[width=0.48\textwidth]{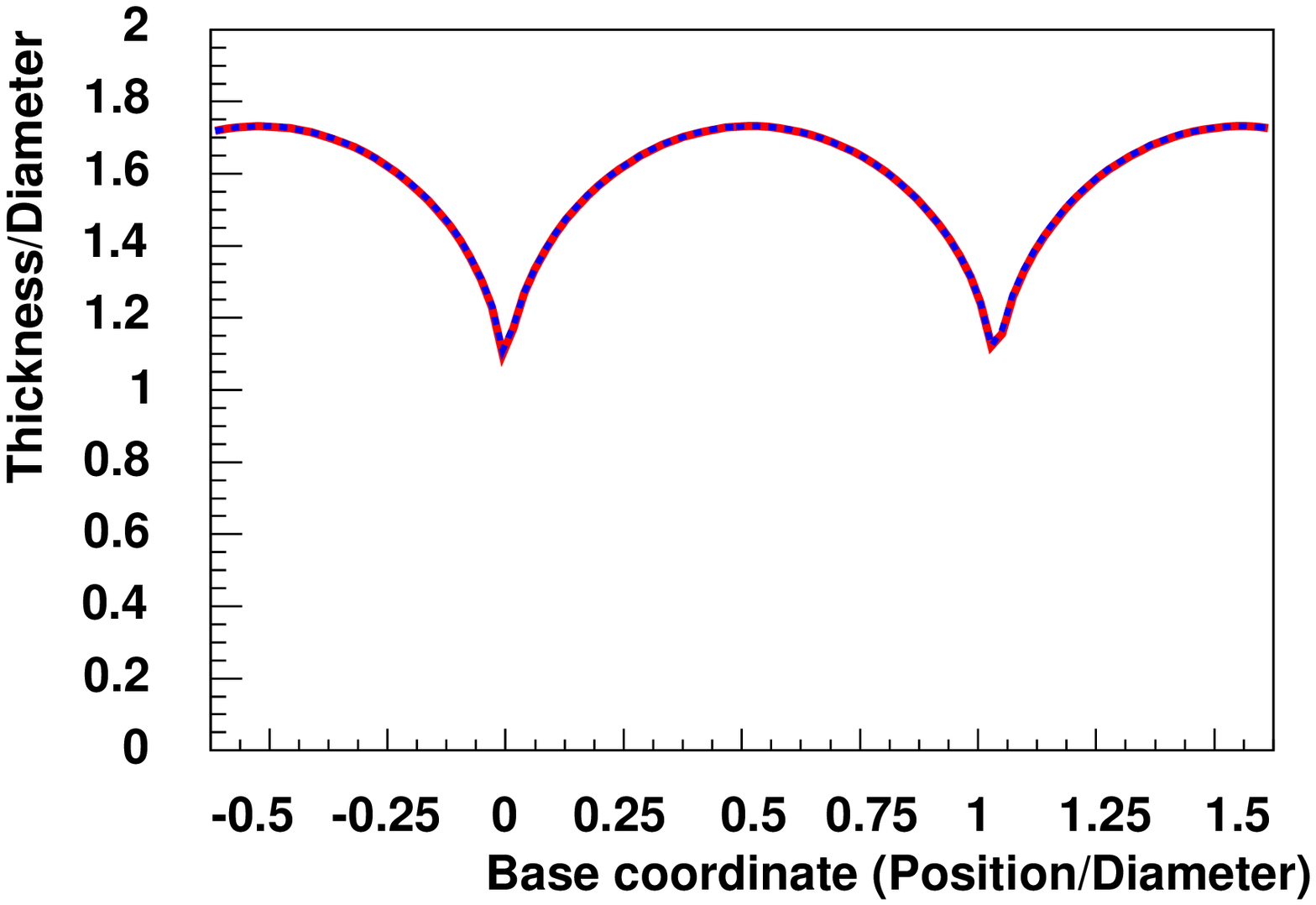}
  \includegraphics[width=0.48\textwidth]{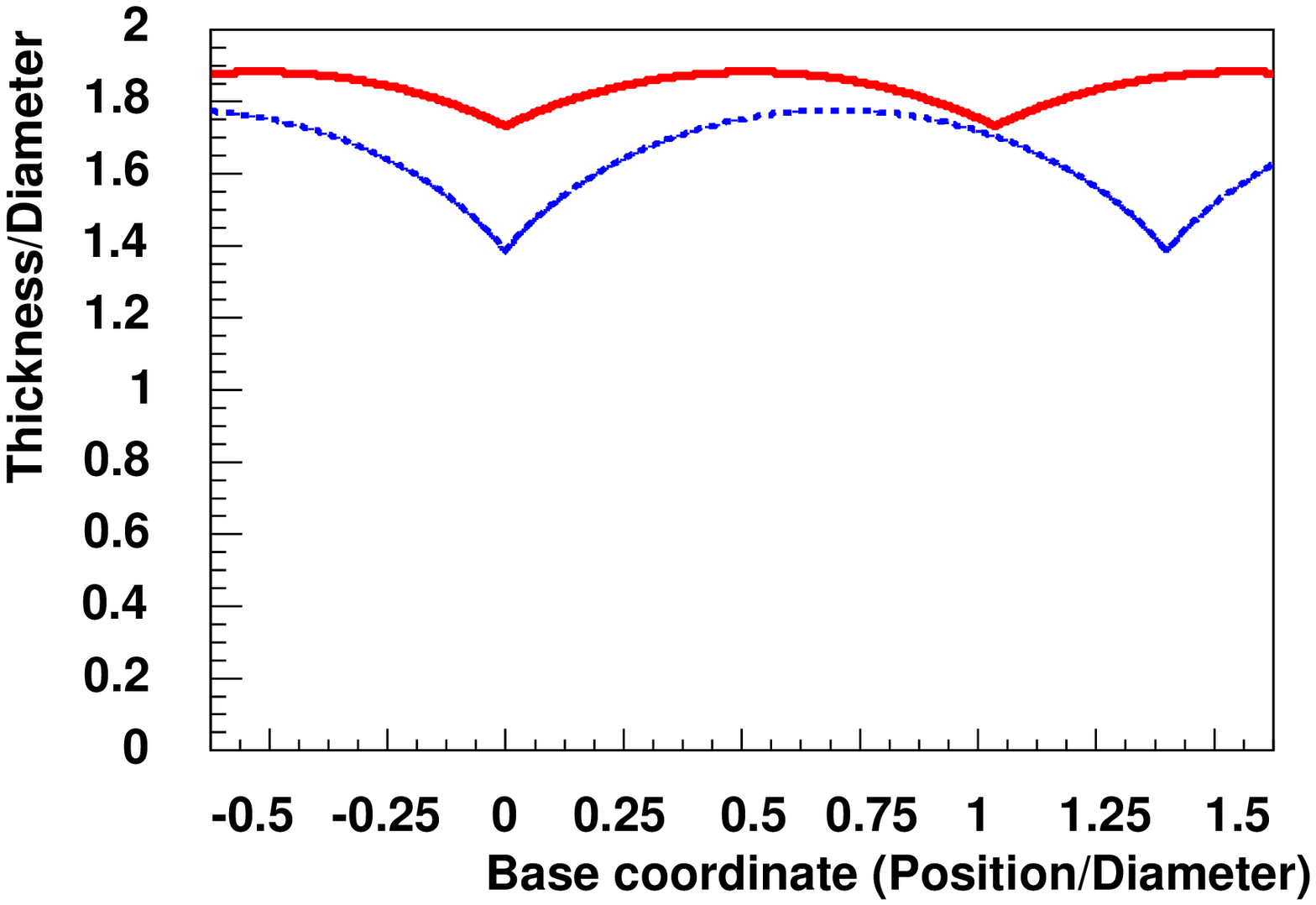}
  \caption{Double layer thickness variation as a function of the base
    coordinate for different fibre array geometries: column angle
    $\phi= 0^\circ$ (top left), $\phi= 45^\circ$ (top right), $\phi=
    60^\circ$ (bottom left), and $\phi= 70^\circ$ (bottom right). The
    two curves correspond to geometries with closed columns (full
    curve) and closed rows (dotted curve).}
  \label{fig:thickness}
\end{figure}

The spread of energy deposition in a fibre bundle is proportional to
the detector thickness variation. Fig.~\ref{fig:thickness} shows how
the double layer thickness varies as a function of the base
coordinate.  In the hexagonal packing with a column angle of $\phi=$
60$^\circ$ the variation is $\delta t\sim 70\,\%\,\oslash$ (minimum to
maximum) per double layer. For all other column angles the variation
can become smaller, depending on the fibre array geometry.

\section{Monte Carlo simulation for diverging particles}
The design criteria discussed so far only apply when all particles
cross the detector with the same incident angle. The expected response
of a detector to diverging particles can be evaluated in a simulation.
Different fibre arrays, in which particles were crossing the detector
with finite incident angles with respect to the column angle, were
included in a detector simulation within the {\sf Geant4}
framework~\cite{GEANT4}. Incident electron trajectories were averaged
over the base. The simulation gave information on the energy
deposition in individual fibres and on interactions of the particles
with the material, e.g.\ small angle scattering, ionisation and
bremsstrahlung.  The total energy deposited in the active cores of the
four corresponding fibres of each column was calculated.  Signals
above a given threshold were assigned to the corresponding read-out
pixel.  

\begin{figure}[htb]
  \centering
  \includegraphics[width=0.48\textwidth]{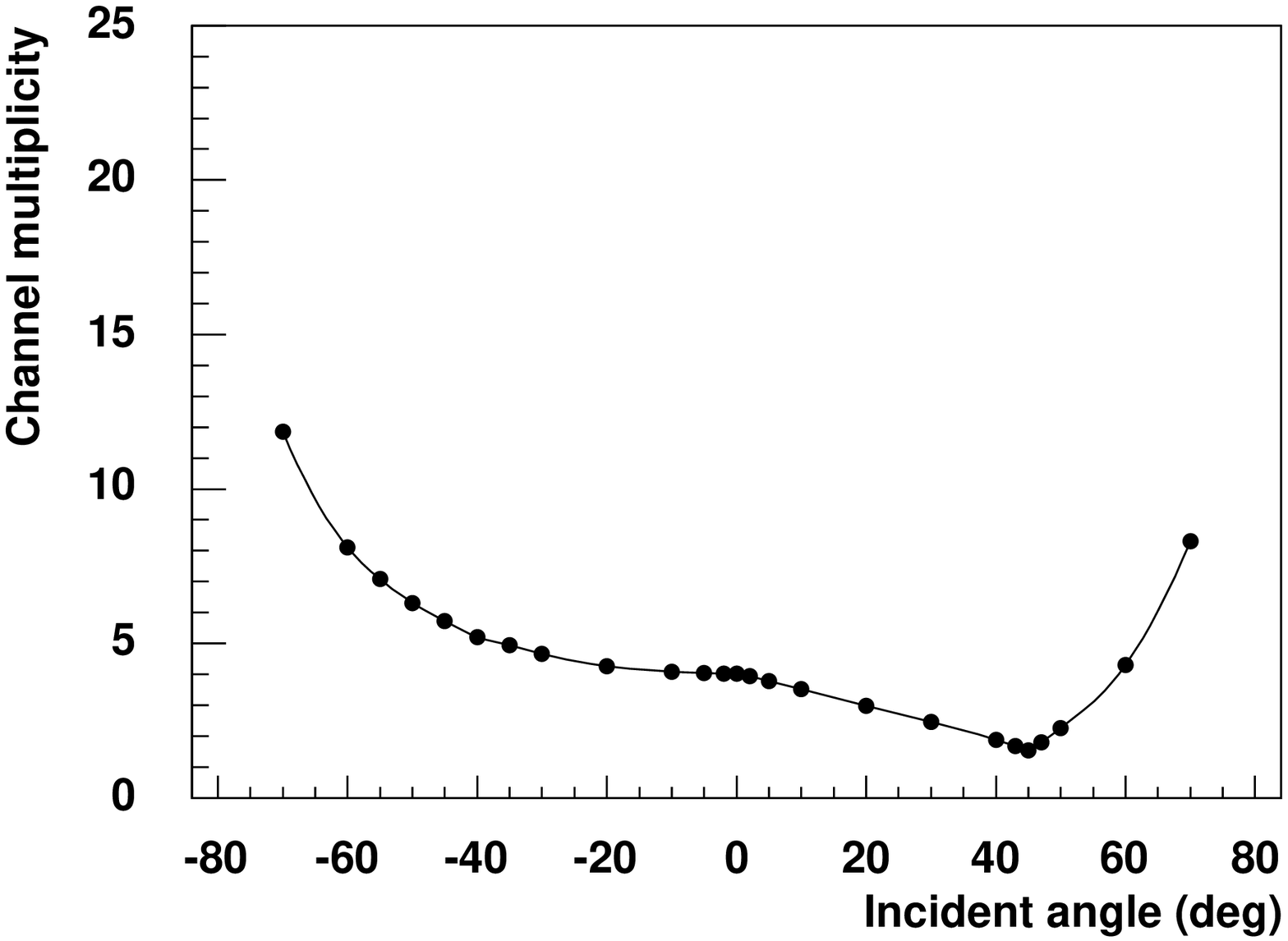}
  \includegraphics[width=0.48\textwidth]{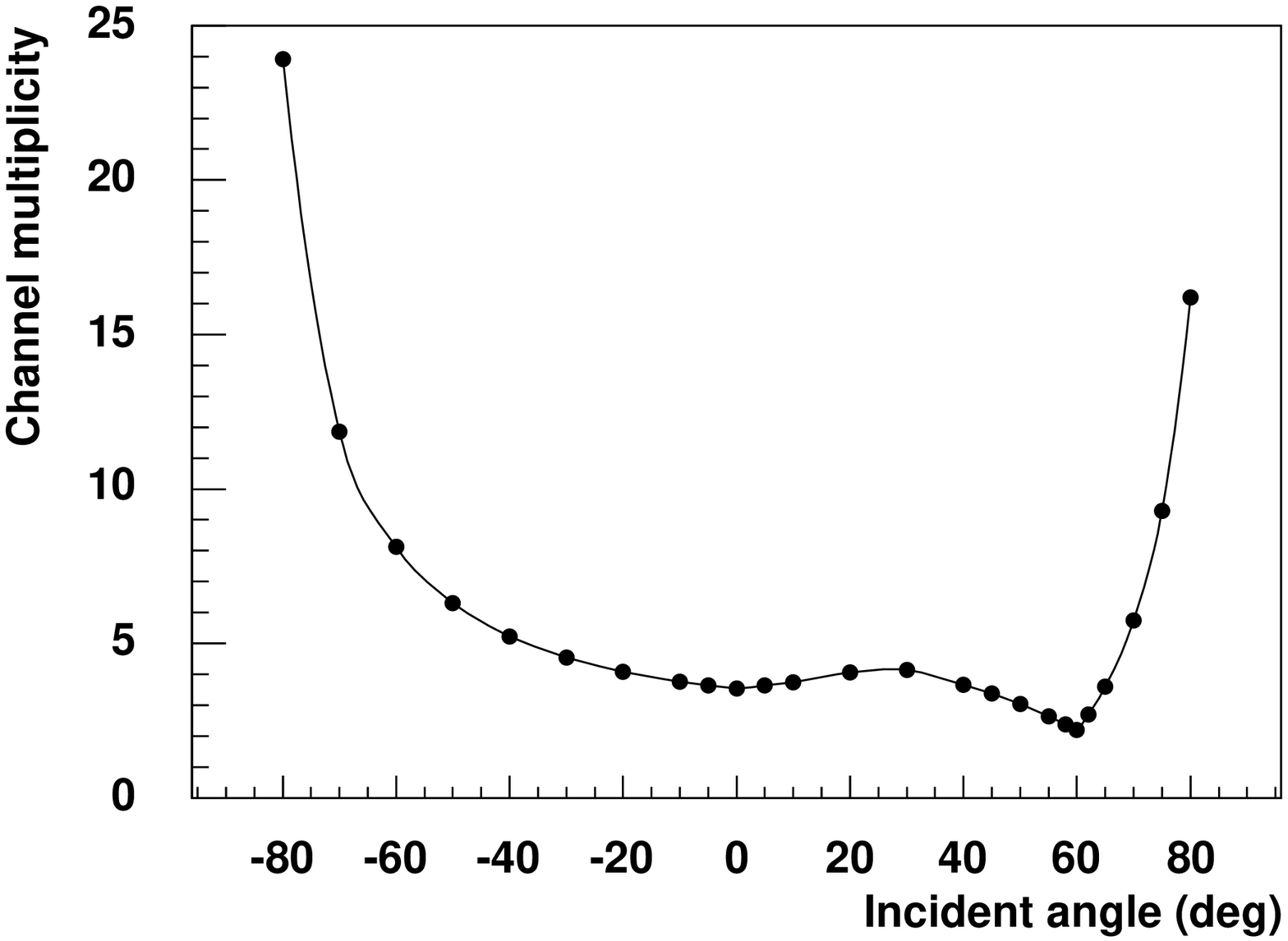}
  \caption{Simulated variation of the channel multiplicity as a
    function of the incident angle for fibre array geometries with
    closed rows and column angles of 45$^\circ$ (left) and 60$^\circ$
    (right).}
  \label{fig:multiplicities}
\end{figure}

The increase in channel multiplicity for detectors with column angles
of $45^\circ$ and $60^\circ$ is presented in
Fig.~\ref{fig:multiplicities} as a function of the incident angle
$\theta$.  It is shown that the multiplicity increases gradually to
smaller incident angles but steeper to higher angles. Owing to the
symmetry of the fibre arrays, a local maximum of the multiplicity
appears for $\phi= 60^\circ$ at $\theta=$ 30$^\circ$, and for $\phi=
45^\circ$ at $\theta=$ 0$^\circ$.  It is obvious that the columns must
be aligned with the incident particle direction and the particle's
divergence must be small to achieve a minimal multiplicity and an
optimal spatial resolution.

\begin{figure}[htb]
  \centering
  \includegraphics[width=0.49\textwidth]{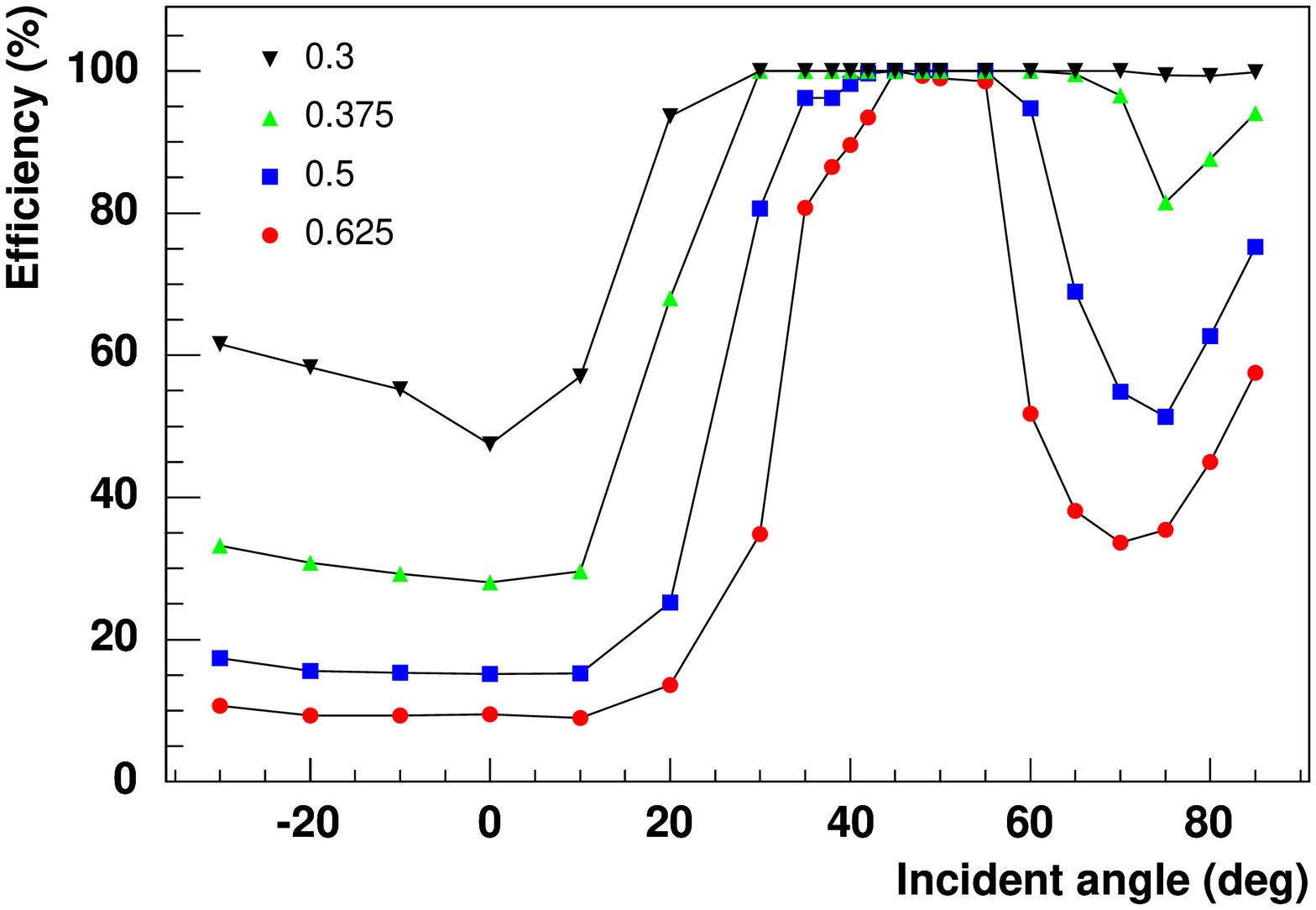}
  \includegraphics[width=0.49\textwidth]{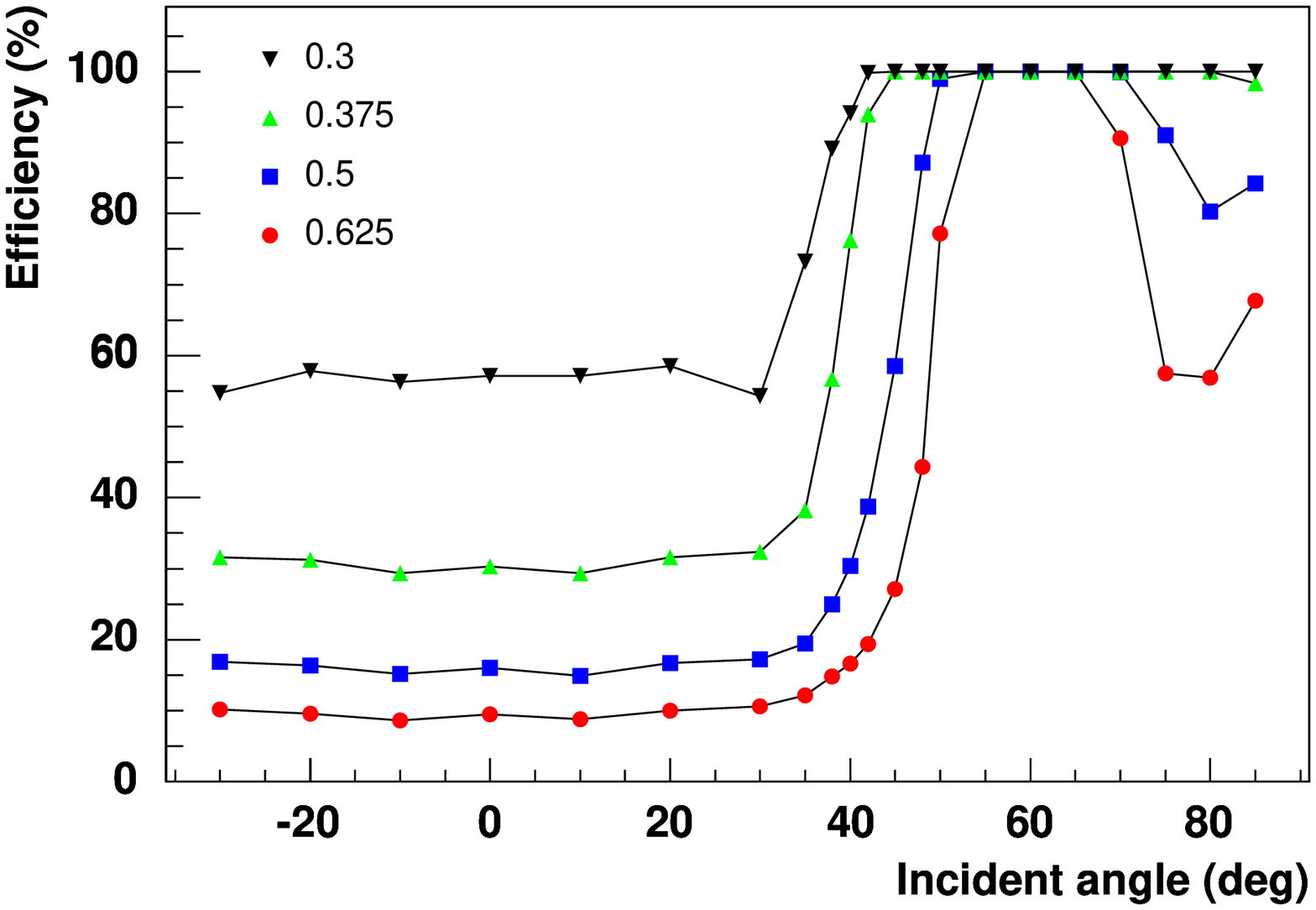}
  \caption{Simulated variation of the detector efficiency as a
    function of the incident angle for fibre array geometries with
    closed rows and column angles of 45$^\circ$ (left) and 60$^\circ$
    (right). Each symbol represents a different threshold relative to
    the mean signal of particles with nominal incident angle.}
  \label{fig:efficiency}
\end{figure}

The simulation can further help to define the optimum threshold so
that a detector becomes insensitive to incoming particles with large
angles w.r.t\ the nominal angle. That is important when a detector is
exposed to a large number of background particles. The angular
dependence of the detector efficiency is plotted in
Fig.~\ref{fig:efficiency} for 4 different relative thresholds.  The
relative threshold of 100\,\% corresponds to the mean signal of a
detector pixel for events with particles hitting the detector with
nominal angle. The plateau width for 100\,\% efficiency depends
crucially on the threshold. For incident angles much smaller than the
nominal angle only one fibre of each column is hit, but many
neighbouring channels.  Thus, the probability for having a signal in
at least one pixel above threshold is strongly dependent on the
distribution of energy deposition and the value of the threshold, but
almost independent on geometry and incident angle.

\section{Summary}
When a set of fibres from a multi-layered fibre detector is coupled to
a multi-channel read-out device a multitude of different fibre array
geometries is possible. It is the experimenter's choice to select the
geometry, which is matching the application best, according to the
relevant design criteria. In this paper the dependence of column pitch
and overlap ratio on the fibre array column angle is discussed.  Both
criteria are also reflected by the thickness variation of a double
layer, which is shown for a selection of column angles. The effect of
diverging particles incident on fibre arrays with inclined columns was
studied with the help of a Monte Carlo simulation. The resulting
curves of channel multiplicity versus incident angle can be folded
with the incident angle distribution of an experiment to get the
average channel multiplicity. Large multiplicities compromise the
position resolution of the fibre arrays, which is the key issue for
tracking detectors. With the choice of the detection threshold, the
sensitivity of the detector to particles with large deviations to the
nominal angle can be varied.

\section*{Acknowledgements}
This work was supported by the Federal State of Rhineland-Palatinate
and by the Deutsche Forschungsgemeinschaft with the Collaborative
Research Center 443.

%%%%%%%%%%%%%%%%%%%%%%%%%%%%%%%%%%%%%%%%%%%%%%%%%%%%%%%%%%%%%%%%%%%%%
%                         BIBLIOGRAPHY                              %
%%%%%%%%%%%%%%%%%%%%%%%%%%%%%%%%%%%%%%%%%%%%%%%%%%%%%%%%%%%%%%%%%%%%%

\clearpage

%%%%%%%%%%%%%%%%%%%%%%%%%%%%%%%%%%%%%%%%%%%%%%%%%%%%%%%%%%%%%%%%%%%%%
%                           FIGURES                                 %
%%%%%%%%%%%%%%%%%%%%%%%%%%%%%%%%%%%%%%%%%%%%%%%%%%%%%%%%%%%%%%%%%%%%%

%
%

%%%%%%%%%%%%%%%%%%%%%%%%%%%%%%%%%%%%%%%%%%%%%%%%%%%%%%%%%%%%%%%%%%%%%
%                             END                                   %
%%%%%%%%%%%%%%%%%%%%%%%%%%%%%%%%%%%%%%%%%%%%%%%%%%%%%%%%%%%%%%%%%%%%%

\end{document}